\documentclass{aastex63}

\usepackage{multirow}

\newcommand{\nicer}{\textit{NICER}}

\newcommand{\swift}{\textit{Swift}}

\newcommand{\xmmlong}{\textit{XMM-Newton}}

\newcommand{\Msun}{\mbox{$M_\odot$}}

\newcommand{\simlt}{\mathrel{\hbox{\rlap{\hbox{\lower4pt\hbox{$\sim$}}}\hbox{$<$}}}}
\newcommand{\simgt}{\mathrel{\hbox{\rlap{\hbox{\lower4pt\hbox{$\sim$}}}\hbox{$>$}}}}

\newcommand{\ud}[2]{\mbox{$^{+ #1}_{- #2}$}}
\newcommand{\ppm}{\mbox{$\pm$}}

\def\deg{\hbox{$^\circ$}}
\def\arcdeg{\hbox{$^\circ$}}
\def\arcmin{\hbox{$^\prime$}}

\newcommand{\G}{\mbox{$\,G$}}
\newcommand{\msun}{\mbox{$\,M_\odot$}}

\newcommand{\K}{\hbox{$\,{\rm K}$}}

\newcommand{\keV}{\mbox{$\,{\rm keV}$}}

\shorttitle{NICER Detection of MSP X-ray Pulsations}
\shortauthors{NICER team}

\begin{document}

\title{NICER Detection of Thermal X-ray Pulsations from the Massive 
Millisecond Pulsars PSR~J0740$+$6620 and PSR~J1614$-$2230}

\correspondingauthor{Michael T. Wolff}
\email{Michael.Wolff@nrl.navy.mil}

\author[0000-0002-4013-5650]{Michael T.~Wolff}
\affil{Space Science Division, U.S. Naval Research Laboratory, Washington, DC 20375-5352, USA}
\author[0000-0002-6449-106X]{Sebastien Guillot}
\affil{IRAP, CNRS, 9 avenue du Colonel Roche, BP 44346, F-31028 Toulouse Cedex 4, France}
\affil{Universit\'{e} de Toulouse, CNES, UPS-OMP, F-31028 Toulouse, France}
\author[0000-0002-9870-2742]{Slavko Bogdanov}
\affil{Columbia Astrophysics Laboratory, Columbia University, 550 West 120th Street, New York, NY 10027, USA}
\author[0000-0002-5297-5278]{Paul S.~Ray}
\affil{Space Science Division, U.S. Naval Research Laboratory, Washington, DC 20375-5352, USA}
\author[0000-0002-0893-4073]{Matthew Kerr}
\affil{Space Science Division, U.S. Naval Research Laboratory, Washington, DC 20375-5352, USA}
\author{Zaven Arzoumanian}
\affil{Astrophysics Science Division, NASA's Goddard Space Flight Center, Greenbelt, MD 20771, USA}
\author{Keith C.~Gendreau}
\affil{Astrophysics Science Division, NASA's Goddard Space Flight Center, Greenbelt, MD 20771, USA}
\author[0000-0002-2666-728X]{M.~Coleman Miller}
\affiliation{Department of Astronomy and Joint Space-Science Institute, University of Maryland, College Park, MD 20742-2421, USA}
\author[0000-0001-6157-6722]{Alexander J.~Dittmann}
\affiliation{Department of Astronomy and Joint Space-Science Institute, University of Maryland, College Park, MD 20742-2421, USA}
\author[0000-0002-6089-6836]{Wynn C.~G. Ho}
\affiliation{Department of Physics and Astronomy, Haverford College, 370 Lancaster Avenue, Haverford, PA 19041, USA}
\author{Lucas Guillemot}
\affiliation{Laboratoire de Physique et Chimie de l'Environnement et de l'Espace, LPC2E, CNRS-Universit\'{e} d'Orl\'{e}ans, F-45071 Orl\'{e}ans, France}
\affiliation{Station de Radioastronomie de Nan\c{c}ay, Observatoire de Paris, CNRS/INSU, F-18330 Nan\c{c}ay, France}
\author{Ismael Cognard}
\affiliation{Laboratoire de Physique et Chimie de l'Environnement et de l'Espace, LPC2E, CNRS-Universit\'{e} d'Orl\'{e}ans, F-45071 Orl\'{e}ans, France}
\affiliation{Station de Radioastronomie de Nan\c{c}ay, Observatoire de Paris, CNRS/INSU, F-18330 Nan\c{c}ay, France}
\author{Gilles Theureau}
\affiliation{Laboratoire de Physique et Chimie de l'Environnement et de l'Espace, LPC2E, CNRS-Universit\'{e} d'Orl\'{e}ans, F-45071 Orl\'{e}ans, France}
\affiliation{Station de Radioastronomie de Nan\c{c}ay, Observatoire de Paris, CNRS/INSU, F-18330 Nan\c{c}ay, France}
\affiliation{LUTH, Observatoire de Paris, PSL Research University, CNRS, Universit\'{e} Paris Diderot, Sorbonne Paris Cit\'{e}, F-92195 Meudon, France}
\author{Kent S. Wood}
\affiliation{TSC, Alexandria, VA, USA} 


\begin{abstract}
We report the detection of X-ray pulsations from the rotation-powered 
millisecond-period pulsars PSR~J0740+6620 and PSR~J1614$-$2230, two 
of the most massive neutron stars known, using observations with 
the \textit{Neutron Star Interior Composition Explorer} (\nicer{}).
We also analyze \textit{X-ray Multi-Mirror Mission} (\textit{XMM-Newton})
data for both pulsars to obtain their time-averaged fluxes and 
study their respective X-ray fields. 
PSR~J0740+6620 exhibits a broad double-peaked profile with a separation 
of $\sim 0.4$ in phase. 
PSR~J1614$-$2230, on the other hand, has a broad single-peak profile. 
The broad modulations with soft X-ray spectra of both pulsars are indicative of 
thermal radiation from one or more small regions of the stellar surface. 
We show the \nicer{} detections of X-ray pulsations for both pulsars and 
also discuss the phase relationship to their radio pulsations.
In the case of PSR~J0740+6620, this paper documents the data reduction 
performed to obtain the pulsation detection and prepare for pulse 
profile modeling analysis.  
\end{abstract}


\section{Introduction}
\label{sec:intro}
Rotation powered millisecond pulsars (MSPs) are usually old neutron 
stars (NSs) with spin periods in the $\sim$1--30 ms range.  
Their fast rotation is thought to result from the transfer of angular 
momentum of accreted matter from a close binary companion during a 
so-called ``recycling'' phase  \citep{alpar82,radhakrishnan82}. 
MSPs can produce emission from radio to gamma-rays 
through particle acceleration and radiation processes in the pulsar magnetosphere. 
However, in the X-ray band substantial emission can originate from
hot spots on the NS surface, and such emission can be an important 
probe of NS properties. 
This paper will focus on the X-ray emission from two 
pulsars found to be massive NSs.

Radio pulse timing observations of MSPs have resulted in precisely 
measured NS masses near or above 2\,M$_{\odot}$ for 
several objects to date.
For PSR~J1614$-$2230, \citet{demorest10} found $M_{\rm NS} = 1.97 \pm 0.04\,M_\odot$ 
and \citet{Arzoumanian_etal2018} found $M_{\rm NS} = 1.908 \pm 0.016\,M_\odot$; 
for PSR~J0348+0432, \citet{antoniadis13} found $M_{\rm NS} = 2.01 \pm 0.04\,M_\odot$; 
for PSR J0740+6620, \citet{cromartie20} found $M_{\rm NS} = 2.14_{-0.09}^{+0.10}\,M_\odot$
and \citet{Fonseca_etal2021} found $M_{\rm NS} = 2.08\pm0.07\,\Msun$.
A potential fourth pulsar with high mass is PSR J1811$-$2405 for which 
\citet{ng_etal2020} find $M_{\rm NS} = 2.0_{-0.5}^{+0.8}\,M_\odot$. 
Finally, radial velocity measurements in the optical band have 
identified potential high NS mass candidates as well. 
\citet{Romani_etal2021} measured $M_{\rm NS} = 2.13 \pm 0.04\,\Msun$ 
for the MSP PSR~J1810$+$1744, and \replaced{\citet{Linares_etal2018} report 
$M_{\rm NS} = 2.27_{-0.15}^{+0.17}\,M_\odot$}{\citet{Kandel2020} report 
$M_{\rm NS} = 2.28_{-0.09}^{+0.10}\,M_\odot$} for the 
redback binary system PSR~J2215$+$5135 \added{\citep[see also][]{Linares_etal2018}}. 
Indeed, \citet{Romani_etal2021} explore the possibility that 
a substantial fraction of the NS population may exist with 
high masses near 2.0 $M_\odot$ and above.
\added{How NSs can acquire such high masses is still an unresolved 
question, however.}
Depending on the evolutionary pathway of a given MSP, an 
appreciable amount of mass can be accreted from the companion 
star: for example, \cite{tauris11} find that PSR~J1614$-$2230 
could have accreted between $\sim$0.05\msun\ and $\sim$0.45\msun. 
\added{On the other hand, \citet{Cognard_etal2017} argue that the moderately
massive NS in PSR~J2222$-$0137 ($M_{\rm NS} = 1.75 \pm 0.06\,M_\odot$) likely 
acquired its mass at birth and not via accretion.}

Such massive NSs are of great interest for understanding the NS 
mass-radius relation and the equation of state (EOS) of cold, dense matter.
High-mass NSs enable us to constrain the very uncertain EOS at the 
highest densities and connect with the EOS at lower densities 
obtained from less massive NSs.
Furthermore, each theoretical model of NS structure predicts 
a particular maximum  
mass (\citealt{oppenheimer39}; see also, e.g., \citealt{lattimer12}), 
above which black hole formation occurs.
For example, GW190814 indicates the merger of a black hole and a second compact 
object with mass 2.6\msun\ \citep{abbott20}.  
If the second object is also a black hole, then the maximum mass of a 
(slowly-rotating) NS is between 2\msun\ and 2.6\msun, and there is either a 
small gap or no gap between the highest mass NS and lowest mass black hole.  
On the other hand, if knowledge gained from 2\msun\ NSs indicates an EOS that 
can support much higher mass NSs, then objects such as the 
secondary of GW190814 could be a NS (see \citealt{Abbott_etal2018a}).

Even more stringent constraints on the dense matter EOS can be obtained 
if the mass measurement is combined with a radius estimate from modeling 
of the pulsed surface thermal X-ray radiation \citep[see, e.g.,][]{pavlov97,miller98b,bogdanov07,Psaltis_and_Ozel2014}. 
Obtaining NS radius measurements using this technique is one of the 
primary scientific goals of the \textit{Neutron star Interior 
Composition Explorer} (\nicer{}) mission. 
The first such constraints were presented in \citet{miller19} 
and \citet{riley19} based on a deep \nicer{} exposure of the nearby 
isolated MSP PSR J0030$+$0451 \citep{bogdanov19a}.

\nicer{} observations of massive MSPs have the potential to provide unique 
constraints on the properties of cold, catalyzed matter beyond nuclear 
density even when very X-ray faint, as discussed in depth 
by \citet{miller16} for the case of PSR~J1614$-$2230.  
Because more compact stars tend to have lower fractional modulation 
amplitudes, all else being equal, measurement of the \added{thermal} 
X-ray pulsation amplitude 
will provide an upper limit on the compactness $GM/Rc^2$ and thus a 
lower limit on the radius given the precisely-determined mass of this star.  
A lower limit to the radius of a high-mass NS, if that lower limit 
exceeds 10--11~km, would provide important input to nuclear theories. 
An upper limit on the stellar radius cannot be obtained in the same way.
For example, moving the spot center toward a rotational pole can reduce the 
amplitude by an arbitrary factor regardless of how large the star is.
A robust minimum radius can be inferred, using assumptions that otherwise 
maximize the overall amplitude (including the assumptions that 
the spot is point like and centered on the rotational equator). 
\added{This is contingent, however, on being able to place constraints 
on any weak non-thermal pulsar emission that could potentially bias the 
derived radius constraints \citep{guillot16a}. }

PSR J0740+6620 is an MSP with $P_{\rm{spin}} = 2.89$\,ms discovered in the 
course of the Green Bank Northern Celestial Cap (GBNCC) pulsar 
survey \citep{lynch2018}. 
It is in a nearly circular 4.8 day binary orbit, possibly with an ultra-cool 
white dwarf companion \citep{beronya2019}.  
Recently, \citet{Fonseca_etal2021} obtained an updated mass measurement 
of $M_{\rm NS} = 2.08\pm0.07$ $M_{\odot}$ (at 68.3\% confidence) for PSR~J0740+6620, 
from the observed radio Shapiro delay, making it the NS with the largest precisely 
established mass (first reported by \citealt{cromartie20}).  
\citet{Fonseca_etal2021} also reported an updated distance measurement from timing parallax and 
the Shklovskii effect of $d = 1.14^{+0.17}_{-0.15}$ kpc (68.3\% confidence).

PSR~J1614$-$2230 was discovered in a Parkes radio search targeting 
unidentified EGRET sources \citep{hessels05,crawford06}. 
It is an MSP with $P_{\rm{spin}} =3.15$\,ms bound to a massive white dwarf 
companion in a $P_{\rm{b}} = 8.7$\,d orbit. 
The pulsar is particularly important for the \nicer{} mission because 
it is one of the most massive NSs known, with the most recent 
measurement of $M=1.908\pm0.016\msun$ \citep{arzoumanian18}. 
This precise measurement resulted from the detection of a strong 
signature of Shapiro delay in the binary \citep{demorest10}. 
PSR~J1614$-$2230 has a parallax of 1.5\ppm0.1\,mas, so it lies 
at a parallax distance of $d=670\ud{50}{40}$\,pc \citep{arzoumanian18}. 
X-ray pulsations from this pulsar were first claimed at 
the $\sim 4\sigma$ confidence level based on \textit{XMM-Newton} 
observations \citep{pancrazi12}.

Initial exploratory X-ray observations of both pulsars with \nicer{}, 
\xmmlong, and \swift{} showed that they are very faint sources.
We have undertaken a systematic survey with \nicer{} of the sample of nearby 
rotation-powered MSPs to detect and characterize their pulsed X-ray radiation, with a 
focus on finding thermal X-ray pulsations that are desirable for constraints on 
the NS mass-radius relation and the dense matter EOS. 
The \nicer{} observations of eight other MSPs were presented in \citet{ray19} 
and \citet{guillot19}. 

In the present paper, we describe \nicer{} observations of the two nearby binary 
MSP systems hosting massive neutron stars, PSR~J0740+6620 and PSR~J1614$-$2230, and 
report the detection of thermal X-ray pulsations from both. 
The paper is organized as follows. 
In Section~\ref{sec:obs} we describe the X-ray 
observations and the data reduction procedures.
Because the photon event data for PSR~J0740$+$6620 have been made
available for the NS radius inference analyses of \citet{Miller_etal2021} and
\citet{Riley_etal2021}, the data extraction procedure is more 
stringent for that pulsar than for PSR~J1614$-$2230.
In Section~\ref{sec:results} we present our findings from 
the X-ray observations together with the known radio properties of these two pulsars.
In Section~\ref{sec:discussion}, we discuss issues that are raised either by
the observations themselves or the follow-on inference analyses.
Finally, in Section~\ref{sec:conclusions} we summarize our principal conclusions. 

\section{Observations}
\label{sec:obs}

The \textit{Neutron Star Interior Composition Explorer} 
\citep[\nicer; see][]{gendreau17} is a NASA astrophysics explorer mission 
that has been operating on the International Space Station (ISS) since 2017 June. 
The \nicer{} X-ray Timing Instrument (XTI) is an array of 52 active silicon 
drift detectors each paired with concentrator optics that are nominally sensitive 
to photons in the 0.2$-$12~keV range and have absolute timing precision 
of $\sim$100 nanoseconds rms for tagging of photon detection times \citep{2016SPIEPrigozhin}.  
The event data for both sources were processed using {\tt HEASOFT} v6.28 and 
the \nicer{}-specific package {\tt NICERDAS} v6, together with the \nicer{} 
calibration files v20200202.
However, somewhat different criteria for 
extracting \nicer{} event data were used for the two pulsars. 
Furthermore, the \xmmlong{} data were obtained under a number of different 
circumstances, as described below.

\begin{deluxetable}{lcccc}
\tablecaption{\nicer\ observation summary for PSR~J0740+6620\label{tab:0740nicer}}
\tablehead{
  & \colhead{ObsID} & \colhead{Dates}  & \colhead{Exposure}  &  \colhead{Exposure} \\
  & \colhead{Range} & \colhead{Range} & \colhead{Raw (ks)}  &  \colhead{Filtered (ks)}
}
\startdata
  & 1031020101 -- 1031020128 & 2018-09-21 -- 2018-10-29 & 322.1  & 302.8 \\
  & 2031020101 -- 2031020257 & 2019-04-05 -- 2020-02-29 & 1142.2 & 983.5 \\
  & 3031020201 -- 3031020244 & 2020-03-01 -- 2020-04-17 & 403.2  & 360.1 \\
\enddata
\end{deluxetable}

\begin{deluxetable}{lcccc}
\tablecaption{\nicer\ observation summary for PSR~J1614$-$2230\label{tab:1614nicer}}
\tablehead{
  & \colhead{ObsID} & \colhead{Dates}  & \colhead{Exposure}  &  \colhead{Exposure} \\
  & \colhead{Range} & \colhead{Range} & \colhead{Raw (ks)}  &  \colhead{Filtered (ks)}
}
\startdata
 &  0060310101 -- 0060310109 & 2017-07-04 -- 2017-07-16 & 17.2  & 15.2 \\
 & 	1060310101 -- 1060310257 & 2017-07-18 -- 2019-03-01 & 410.3 & 332.3 \\
 & 	2060310201 -- 2060310264 & 2019-03-06 -- 2019-09-12 & 143.7 & 126.8 \\
 & 	3060310201 -- 3060310298 & 2020-04-02 -- 2021-03-01 & 281.3 & 260.0 \\
		 \enddata
\end{deluxetable}

\subsection{NICER}
\label{sec:nicerobs}

PSR~J0740+6620 was observed by \nicer{} for 1867.5~ks between 2018 September 21 and 
2020 April 17, while PSR~J1614$-$2230 was observed by \nicer{} for 852.5~ks over the 
period spanning 2017 July 4 to 2021 March 1. 
Both pulsars are challenging to observe because of their faintness,
although PSR~J1614$-$2230 is somewhat brighter than PSR~J0740+6620. 
Because both are faint, everything possible has been done to minimize 
backgrounds, including making observations primarily during ISS orbital night. 
Special care is warranted for PSR~J0740+6620 because data sets are being 
used to obtain a NS radius constraint from the light curve modeling of the 
pulsed component (which is derived from \textit{NICER}) relative to the total 
phase-averaged flux (which is derived from \xmmlong{} imaging).  
Furthermore, there are nearby sources of comparable brightness that the \nicer{}
concentrating detector optics do not exclude and their contributions must be estimated.
A summary of \nicer{} observations of our target pulsars can be found in Tables 
\ref{tab:0740nicer} and \ref{tab:1614nicer}.

For the pulsar PSR~J0740+6620, all events included in our and subsequent 
analyses were obtained when the pulsar was at least 80\deg{} from the Sun. 
This minimized the effects of scattered solar X-rays and detector 
noise resulting from photons whose energies are just below the 
NICER energy range. 
We extract the event data in an ObsID by ObsID manner.
We first filter out event data outside the 0.25$-$3.0\keV\ energy range. 
We also filter the observations keeping only those times that have a Cosmic Ray 
Cutoff Rigidity greater than 2.0 GeV/$c$ and a space weather Kp index less than 5.0.
We further require that the observations by \nicer{} include the full 
complement of 52 detectors active during the observations. 
We exclude the detector with \texttt{DET\_ID} 34 since this detector tends to be noisy.
This ensures that the event data passed to the light curve analysts  
have an internally consistent set of 51 out of 52 available detectors and 
thus utilize a constant effective area configuration.

We also demand that \textit{NICER} is always pointed to better than
15 arcseconds of the source in order to include the data. 
The \textit{NICER} vignetting function is known to be relatively flat-topped
within 2 arcminutes of the instrument boresight so the pointing accuracy filter
removes rare, large pointing errors as a possible significant source of brightness variation.
In August 2020 maintenance on the individual detectors was begun and this involved
cycling individual detectors off for a period of annealing. 
Thus, for our initial investigation of PSR~J0740+6620, utilizing event data obtained 
after the detector maintenance was started would have created a \nicer{} XTI
effective area inconsistency across this date boundary and we 
choose to avoid this complication.
The pulsar phase for each event is calculated with the \texttt{photonphase} 
tool of the \texttt{PINT} software package \citep{pint} using the timing 
solutions for each pulsar described in \S \ref{sec:radio}.  
The end result here is a set of good time intervals (GTIs) with event data 
and low background noise. 

For PSR~J1614$-$2230, because we are not yet at the stage of performing a full light curve analysis to
obtain an estimated NS radius, we slightly relaxed our event acceptance criteria.
Namely, we employ a Kp index constraint of 4.0 and a Sun angle constraint of 60\deg{}.
Furthermore, some of the \nicer{} detectors suffered from a brief single event 
upset (SEU) on 8 July 2019, that caused them to develop anomalous time stamps until 
those detectors could be reset.
We exclude the three days of \nicer{} event data between the SEU occurrence and the reset.
Because PSR~J1614$-$2230 is somewhat brighter than PSR~J0740+6620 this change in event 
acceptance criteria was found to be the best trade-off between integration time
accumulation and background noise.

The \nicer{} non-imaging field of view (FOV) with a $\sim 3\arcmin$ radius 
does not resolve X-ray sources near the targets and thus such sources can
contribute to the effective background counts for the target source.  
Therefore, the true X-ray phase-averaged count rate for faint pulsars may be 
difficult to accurately determine.
In order to partially mitigate this issue, we make use of \xmmlong{} observations 
to better determine the X-ray fluxes of the pulsars and investigate the 
presence of other X-ray sources in the \nicer{} FOV.

\subsection{XMM-Newton}
\label{sec:xmmobs}

\begin{deluxetable}{lcccc}
\tablecaption{\xmmlong\ observation summary for PSR~J0740+6620\label{tab:0740xmm}}
\tablehead{
\colhead{Instrument} &\colhead{ObsID} &\colhead{Date} &\colhead{Exposure} & \colhead{Exposure} \\
                     &                & \colhead{}    & \colhead{Raw (ks)} & \colhead{Filtered (ks)} 
}
\startdata
XMM-MOS1 & 0851181601 & 2019-10-26 &  6.55 & 5.74 \\
	  	 & 0851181401 & 2019-10-28 &  8.69 & 4.94 \\
	  	 & 0851181501 & 2019-11-01 & 11.45 & 7.28 \\
\hline
XMM-MOS2 & 0851181601 & 2019-10-26 &  9.33 & 7.49 \\
	  	 & 0851181401 & 2019-10-28 &  8.68 & 3.33 \\
	  	 & 0851181501 & 2019-11-01 & 11.44 & 7.86 \\
\hline
XMM-pn   & 0851181601 & 2019-10-26 &  6.42 & 4.11 \\
	  	 & 0851181401 & 2019-10-28 &  6.00 & 0.91 \\
	  	 & 0851181501 & 2019-11-01 &  8.42 & 1.78 \\
\enddata
\tablecomments{All \xmmlong\ exposures were obtained in `Full-frame' mode. 
The ``Raw'' exposure times report the value of the ``LIVETIME'' FITS 
keyword. The ``Filtered'' exposure times report the sum of the GTIs 
after the filtering described in Section~\ref{sec:xmmobs}.} 
\end{deluxetable}

\begin{deluxetable}{lcccc}
\tablecaption{\xmmlong{} observation summary for PSR~J1614$-$2230\label{tab:1614xmm}}
\tablehead{
\colhead{Instrument} &\colhead{ObsID} &\colhead{Date} &\colhead{Exposure} & \colhead{Exposure} \\
                     &                & \colhead{}    & \colhead{Raw (ks)} & \colhead{Filtered (ks)} 
}
\startdata
XMM-MOS1 & 0404790101 & 2007-02-08 & 45.6 & 37.3  \\
\hline
XMM-MOS2 & 0404790101 & 2007-02-08 & 45.7 & 37.0  \\
\hline
XMM-pn   & 0404790101 & 2007-02-08 & 38.7 & 21.9  \\
\enddata
\tablecomments{All \xmmlong\ exposures were obtained in `Full-frame' mode. 
The ``Raw'' exposure times report the value of the ``LIVETIME'' FITS 
keyword. The ``Filtered'' exposure times report the sum of the GTIs 
after the filtering described in Section~\ref{sec:xmmobs}.} 
\end{deluxetable}

For PSR~J0740+6620, we acquired three separate \xmmlong{} data sets on 2019 
October 26 (ObsID 0851181601), 2019 October 28 (ObsID 0851181401), and 2019 
November 1 (ObsID 0851181501) through the Director's Discretionary Time program. 
The three EPIC instruments were operated in `Full Frame' mode with 
the `Thin' optical blocking filters in place.  
Due to the long instrument  event readout times in `Full Frame' mode (73.4~ms for 
EPIC-pn and 2.6~s for EPIC-MOS), the data cannot be used for a pulse timing analysis.  
As these observations occurred close to the telescope perigee passage, a large fraction 
of the exposures were affected by intense particle flaring. 
These are filtered out using the standard procedure, with special care to 
exclude times of high background flaring. 
For the EPIC-MOS cameras, we extracted events with  \texttt{PATTERN<=12}, and for
the EPIC-pn camera we extracted events with \texttt{PATTERN<=4}. 
For both instruments we excluded time ranges with significantly higher than average 
count rates in both the full 0.2--12 keV band and the high energy bands 
(10--12 keV for EPIC-MOS and 10--15 keV for EPIC-pn).
This procedure resulted in the exposure times listed in Table~\ref{tab:0740xmm}.  

For PSR~J1614$-$2230, there are four separate observations of the pulsar in the 
\xmmlong{} HEASARC archive. 
We utilize archival \xmmlong{} observations (ObsID: 0404790101) originally 
reported by \citet{pancrazi12}.

\subsection{Radio pulsar timing}
\label{sec:radio}

To search for pulsations, we require pulsar timing models that provide sufficient 
precision over the full \textit{NICER} observation span to assign pulse phases to the 
\textit{NICER} data without needing any additional trials over folding periods. 
These models are provided by ground-based pulsar timing programs at radio frequencies.

PSR~J0740+6620 is regularly observed as part of the NANOGrav pulsar timing 
program \citep{NG12} with both the Green Bank Telescope (GBT) and Canadian 
Hydrogen Intensity Mapping Experiment (CHIME). 
\citet{Fonseca_etal2021} have performed an analysis of these data, extending 
the work done by \citet{cromartie20}. 
Since this covers the entire span of the \textit{NICER} data, we use that timing 
model to assign pulse phases for the present work.

For PSR J1614$-$2230, we constructed a timing model using a merged dataset from 
the NANOGrav 12.5yr release \citep{NG12} and the Nan\c{c}ay radio telescope (NRT). 
The NRT data were obtained from MJD 57424 (2016 Feb 6) through 58921 (2020 Mar 13). 
The NRT backends and data processing procedure are described in \citet{guillemot16}. 
We analyzed this combined data set using \textsc{Tempo2} and fit a pulsar timing model 
containing astrometric, rotational, and binary parameters for the pulsar. 
The model included a time-variable dispersion measure (DM) using a second-order 
polynomial plus a solar wind model. 
The resulting  pulse timing solution is presented in Table \ref{tab:ephem1614}.

\begin{deluxetable}{lr}[ht]
\tablecaption{Ephemeris of PSR~J1614$-$2230.  Digits in 
parentheses represent the $1\sigma$ uncertainty on the 
last quoted digit of a parameter value. \label{tab:ephem1614}}
\tablehead{ \colhead{Parameter} & \colhead{Value} }
\startdata
Right Ascension (J2000) \dotfill 				                            & 16:14:36.507995(7) \\
Declination (J2000) \dotfill 					                            & $-22$:30:31.3329(5)  \\
Proper motion in R.A. (mas yr$^{-1}$) \dotfill 			                    & 3.87(3) \\ 
Proper motion in Decl. (mas yr$^{-1}$) \dotfill			                    & $-32.3(2)$ \\
Parallax, $\pi$ (mas)\dotfill & 1.58(4) \\ 
Epoch of position (MJD)\dotfill   					                            & 56823 \\
\hline
Spin frequency, $\nu$ (Hz) \dotfill				                            & 317.3789418919128(3)  \\ 
Spin frequency derivative, $\dot{\nu}$ (s$^{-2}$)\dotfill 	                & $-9.69490(4) \times 10^{-16} $ \\
Epoch of period (MJD)\dotfill 						                            & 56823 \\
Dispersion measure, DM (pc\,cm$^{-3}$) \dotfill 		                    & 34.48759(3) \\ 
Dispersion measure derivative, DM1 \dotfill 			                    & $-0.000302(6)$ \\
Dispersion measure 2nd derivative\tablenotemark{a}, DM2 \dotfill 			                    & $-1.75(4) \times 10^{-4}$\\
Epoch of DM (MJD)\dotfill   							                            & 56823 \\
Solar wind density at 1 AU (NE\_SW; cm$^{-3}$)\dotfill & 7.33(16) \\ 
\hline
Binary model \dotfill 							                            & ELL1 \\
Binary orbital period, $P_{\rm{b}}$ (days)\dotfill 		                    & 8.6866194173(2)   \\ 
First derivative of orbital period, $\dot{P_b}$\dotfill & 1.28(5)$\times 10^{-12}$ \\ 
Projected semi-major axis of orbit, $x$ (lt-s)\dotfill 		                & 11.29119744(4) \\ 
Epoch of ascending node passage, T$_{\rm asc}$ (MJD) \dotfill 	            & 52331.17011020(5)  \\ 
First Laplace parameter, $\sin\,\omega$ \dotfill   		                    &  $9.7(6) \times 10^{-8}$ \\
Second Laplace parameter $\cos\,\omega$ \dotfill   		                    &  $-1.345(4) \times 10^{-6}$\\ 
Companion mass, $M_c$ ($M_\odot$)\dotfill & 0.4943(19) \\ 
Sine of orbital inclination, $\sin i$ \dotfill                              & 0.999896(4) \\
\hline
Terrestrial time standard (CLK)  \dotfill      			                    & TT(BIPM2017) \\
Units of barycentric time  \dotfill      			                    & TDB \\
Solar System ephemeris  \dotfill						    	            & DE421 \\
\enddata
\tablenotetext{a}{The default definition of DM2 (and higher terms) in \textsc{Tempo2} 
changed in 2020 June. 
We use the corrected Taylor series representation here.}
\end{deluxetable}

\section{Pulsation Detection}
\label{sec:results}

\begin{figure*}[b]
\begin{center}
\gridline{\fig{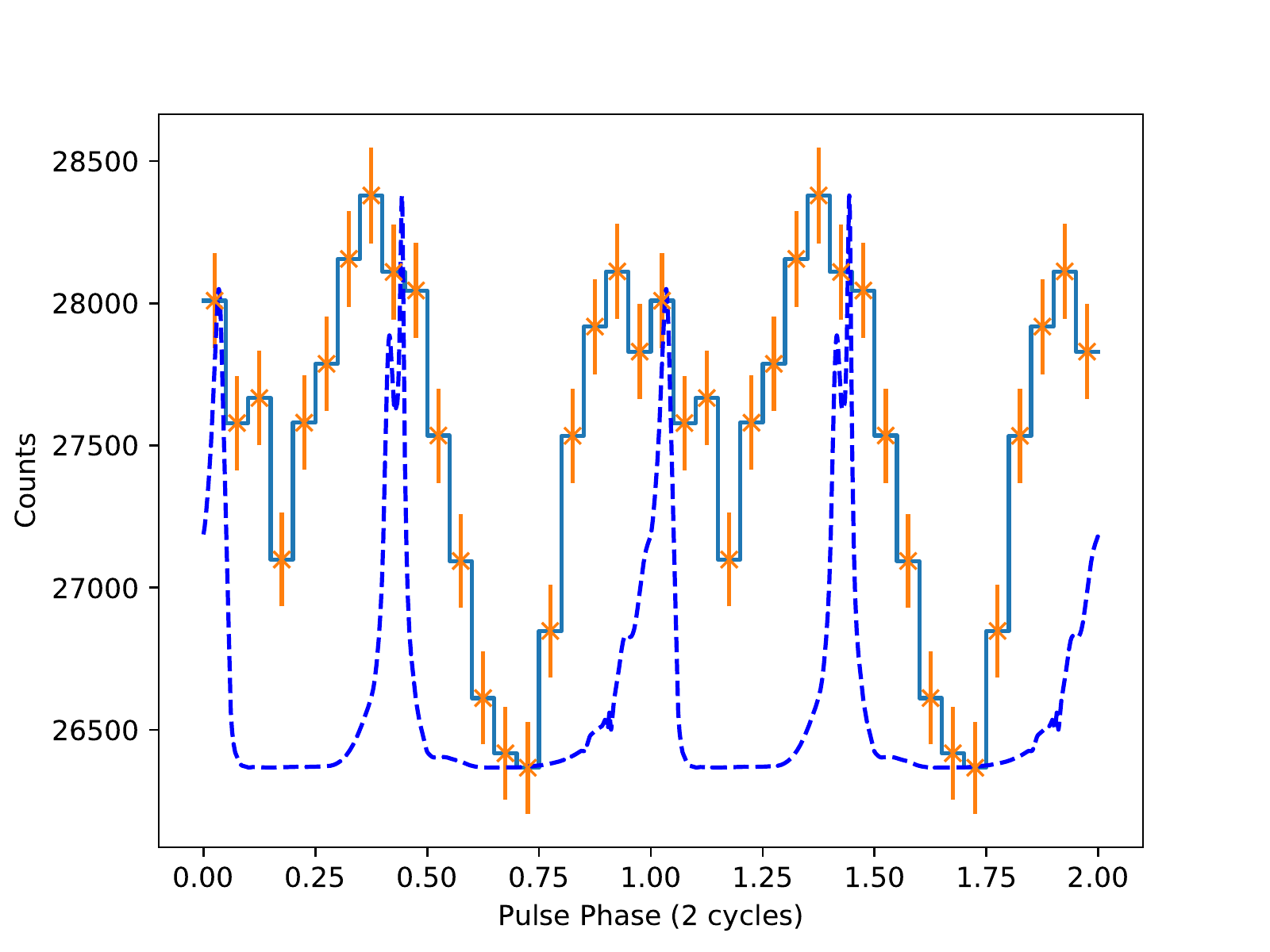}{0.49\textwidth}{(a)}
          \fig{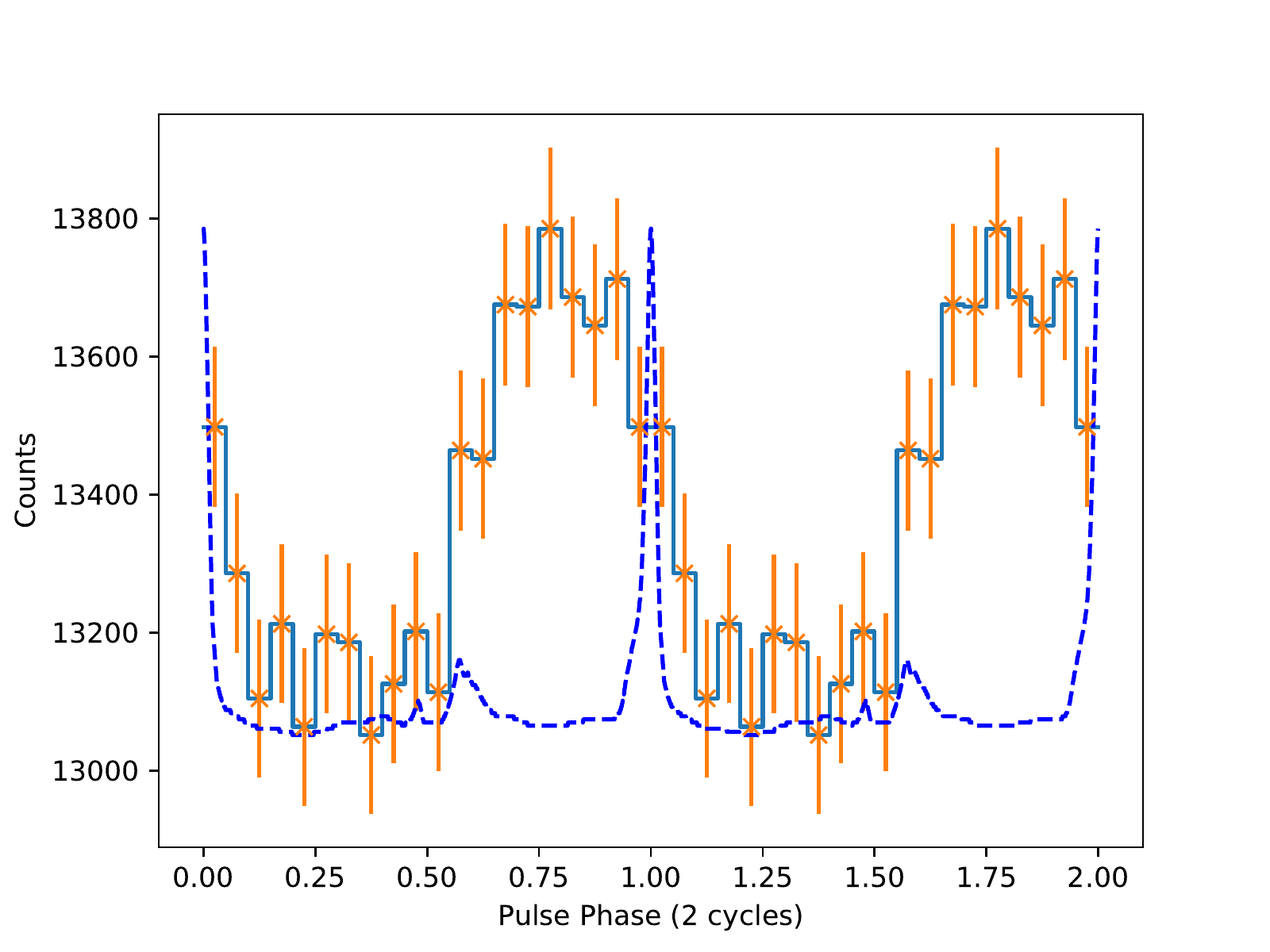}{0.49\textwidth}{(b)}}
\caption{\nicer{} pulse profiles of (a) PSR~J0740+6620 in the 
energy range 0.30--1.25 keV\, and of (b) PSR~J1614$-$2230 in the energy 
range 0.59--2.20 keV, folded with their respective radio spin ephemerides.  
Two rotational cycles are shown for clarity. 
The blue dashed lines correspond to the radio profiles obtained with the NRT 
in the 1.4 GHz band.}
\label{fig:0740_1614_radio}
\end{center}
\end{figure*}

\begin{figure}[ht]
\begin{center}
\gridline{\fig{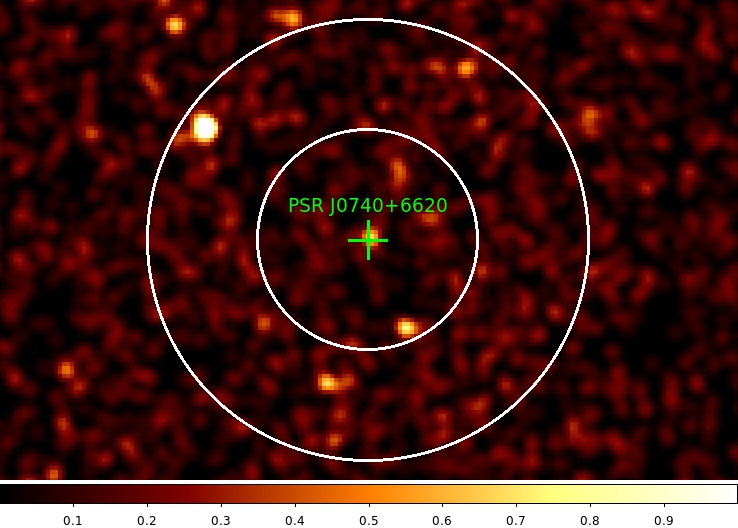}{0.49\textwidth}{(a)}
          \fig{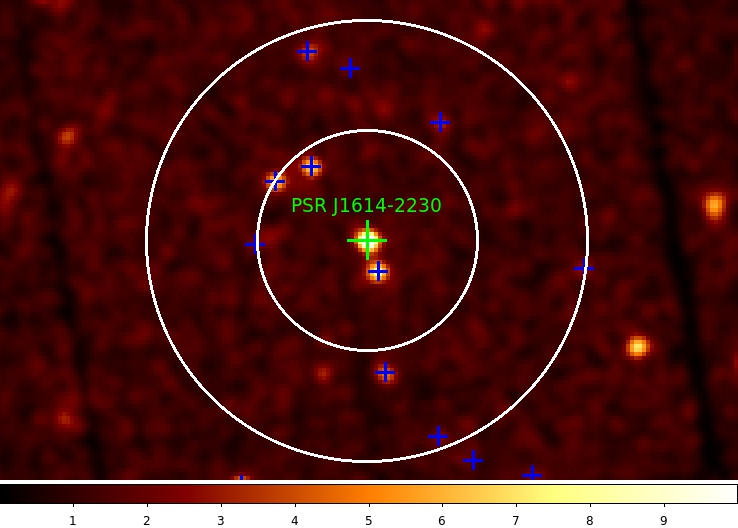}{0.49\textwidth}{(b)}}
\caption{\small{
(Left) \textit{XMM-Newton} combined MOS1 and MOS2 0.5$-$8.0 keV 
image of PSR~J0740$+$6620 and nearby sources.
The green cross shows the radio position of the pulsar and the inner and 
outer white circles are 2 and 4 arcminutes away from the pulsar where the \textit{NICER} 
vignetting function drops quickly from 0.90 to 0.11, respectively. 
The AGN source SDSS J074115.14+662234.9 is 3.57 arcminutes to the north-east of PSR~J0740$+$6620.
At this position SDSS J074115.14+662234.9 falls at the 21\% position 
on the \textit{NICER} vignetting function.
(Right) \textit{XMM-Newton} combined MOS1 and MOS2 0.5$-$8.0 keV 
image of PSR~J1614$-$2230 and nearby sources. 
The green cross shows the radio position of the pulsar and the inner and 
outer white circles are 2 and 4 arcminutes away from the pulsar, respectively. 
The blue crosses are nearby sources from the \textit{Chandra} source catalog 
\citep{Evans_etal2010}.}}
\label{fig:J1614xmm}
\end{center}
\end{figure}

After standard processing and filtering of the \textit{NICER} data as described above, 
we isolate the events from the 0.25 to 3.0 keV range and search for 
pulsations.
Because for both pulsars we have established radio ephemerides we ignore
trials factors and simply assign pulse phases as described above.
For PSR~J0740+6620 a simple Z$_2^2$ test yields a test statistic of
Z$_2^2 = 251.19$, corresponding to a detection 
significance of $15.35 \sigma$ for 574405 events in 
the 0.31 to 1.22 keV range. 
We take advantage of the fact that the filtered \textit{NICER} count rate (about 0.6--1.0 s$^{-1}$) is dominated 
by the background, with the pulsar contributing only a few percent 
to the total count rate. 
We can use the total count rate in each GTI interval as a proxy for 
the effective background count rate and employ the GTI sorting 
method described in \citet{guillot19}. 
The GTI sorting method breaks the original GTI intervals into shorter
segments and adds each segment into the evaluated time series, making 
sure that the overall pulsation significance increases with each addition. 
The lowest count rate segments are included first, and we add in segments 
that have increasing count rates (thus higher background)  so long as 
the detection significance continues to increase. 
Once the background noise becomes sufficiently large that 
adding additional GTIs from the original distribution decreases rather
than increases the detection significance, we stop the procedure.
When we apply this method to PSR~J0740+6620 we increase the overall 
pulsation detection significance to $\sigma = 15.46$ for 521004 
events in the 0.31 to 1.18 keV range. 
The GTI selection is particularly effective for faint pulsars with count 
rates $\sim 0.01$~c/s such as PSR~J0740+6620.
The event lists and GTIs are then saved in both text and FITS file format so that
they can be utilized in subsequent processing, in particular the light curve
modeling of \citet{Miller_etal2021} and \citet{Riley_etal2021} .

For PSR~J1614$-$2230 we confirm the pulsations originally found 
by \citet{pancrazi12} using \xmmlong{} timing-mode data.
Searching the originally extracted 734.3 ks of event data for this 
pulsar in the 0.25 to 3.00 keV range detects the pulsations with 
significance Z$_2^2 = 72.89$ giving $\sigma = 7.81$ for 335614 
events in the 0.59 to 2.20 keV range. 
When we optimize the good time intervals as above, our 
detection significance rises to $\sigma = 8.31$ but the
total time is cut somewhat to 644.9 ks, the number of
included events is likewise reduced to 267122, but the 
optimal energy range is the same at 0.59 to 2.20 keV. 

Figure~\ref{fig:0740_1614_radio} shows the folded \nicer{} pulse 
profiles of PSRs~J0740$+$6620 and J1614$-$2230 in the energy bands 
where the maximum pulse detection significance is found as determined 
by the Z$_2^2$-test \citep{dejager1989,dejager2010}, together with their radio pulse profiles aligned in rotational phase according to the absolute pulse arrival times provided by the radio timing model of each pulsar. 

\section{Discussion}
\label{sec:discussion}

The \nicer{} observations presented here reveal PSR~J0740$+$6620 and 
PSR~J1614$-$2230 as pulsed millisecond-variable X-ray sources with 
phase-broadened thermal pulsations. 
Both NSs have masses near 2~\Msun{} and thus are important 
X-ray sources to elucidate the cold matter EOS near the maximum NS mass.
PSR~J0740$+$6620 shows a double pulse light curve in the \nicer{} energy 
range while PSR~J1614$-$2230 shows one broad pulse. 
The relative phasing of the radio and X-rays shows that for PSR~0740+6620 the
radio pulses are aligned with the thermal X-ray pulses and that the
X-ray pulses are separated by roughly 0.4 in phase.
In the case of PSR~J1614-2230 the dominant radio pulse occurs near the end of the
broad single thermal pulse although there is a very small second pulse in the
radio light curve.
Both pulsars are established as emitting from radio to $\gamma$-ray energies,
indicating that non-thermal emission from the magnetosphere is present.
We found no evidence, however, of non-thermal X-ray emission above 2.5 keV, 
almost certainly because such emission is too faint to be detected in these pulsars.

The goals of these \textit{NICER} and \textit{XMM-Newton} observations are
to establish the existence of X-ray pulsations from both pulsars and 
to estimate the time-averaged flux of each pulsar. 
The \xmmlong{} measurements help constrain the total flux from the NS, enabling 
a better estimate of the pulsed emission fraction in the \textit{NICER} data, 
and thus the estimate of the NS compactness.
A correct pulsed \textit{NICER} flux from the pulsar, if paired with an 
erroneously high pulse-averaged flux from \textit{XMM-Newton}, 
would yield a decreased pulsed fraction and thus an incorrect lower NS radius.
Conversely, a spuriously low \textit{XMM-Newton} flux would yield a 
larger NS radius. 
Because of the high NS masses in these systems, even a solid lower 
limit in the radius of each pulsar would be valuable for constraining the NS
EOS because the cores of such massive NSs contain higher-density matter. 
An upper limit on the compactness, along with the known NS masses, can yield 
a lower limit on the pulsar radius.

These \textit{NICER} data sets for PSR~J0740$+$6620 and PSR~J1614-2230 
are very quiet in that the background count rates are strongly minimized.
Also, for PSR~J0740$+$6620 the effective area as a function of time should 
be very constant because we always utilized the events 
from 51 out of 52 active detectors.

The \xmmlong{} observations of the field of PSR~J0740$+$6620 allow us to
estimate the expected count rates of the likely AGN in the field that
is 3.57 arcminutes north and east from the pulsar. 
In fact, investigation of this X-ray point source shows that there are
really two extragalactic sources within 3.157 arcseconds of one another
[SDSS J074114.62+662235.2 and SDSS J074115.14+662234.9, from the 
Sloan Digital Sky Survey Data Release 9 \citep{Ahn_etal2012}] 
so the \xmmlong{} observations we obtained can not easily distinguish if
the X-rays are coming from one or both of these extragalactic sources.
Using the SAS utility {\tt eregionanalyse} we estimate a \added{0.5$-$3.0 keV 
pulsar count rate of $\sim$0.021 c/s for the \xmmlong{} MOS1.
For an assumed thermal black body pulsar spectrum 
with kT = 0.2 keV and converting the observed radio average dispersion 
measure into an estimate of the absorbing column  
$n_H = 4.5 \times 10^{20}$ cm$^{-2}$ \citep{He_etal2013}, 
the MOS1 count rate translates into a \textit{NICER} count rate of 0.011 c/s in the
0.5$-$3.0 keV energy range.}
\deleted{We assumed here a simple thermal black body spectrum for the pulsar 
with kT = 0.2 keV and converted the observed average dispersion 
measure from the radio observations into an estimate of the 
absorbing column at $n_H = 4.5 \times 10^{20}$ cm$^{-2}$ \citep{He_etal2013}.}
For the position of SDSS J074115.14+662234.9 {\tt eregionanalyse} 
gives a \added{\textit{NICER}} count rate
of $\sim$3.1$\times 10^{-2}$ for this AGN and we assume a non-thermal 
spectrum with power law spectral index $\Gamma \sim 2.71$.
However, since the AGN (at 3.57 arcminutes away) is at the 21\% level
of the \textit{NICER} vignetting curve \citep{takahashi13} as a function 
of angular radius from the look direction of the XTI, this implies the AGN will
only \replaced{give}{add} $\sim 0.7$ \added{of the pulsar count rate} 
\deleted{or a little more than half of the pulsar count rate} 
when \textit{NICER} is pointed directly at PSR~J0740$+$6620.
Weak sources in the field will also contribute some counts because the
\textit{NICER} concentrator optics will cause some photons from these
sources to fall sufficiently close to the center of the detectors to be included 
in the background. 

\section{Conclusions}
\label{sec:conclusions}

The event data from \textit{NICER} and \xmmlong{} we described here 
for PSR~J0740+6620 were used 
by \citet{Miller_etal2021,Raaijmakers_etal2021,Riley_etal2021}
to investigate the radius and EOS of the pulsar in this system.
The event list we derive for this pulsar is available for download 
on Zenodo.org website\footnote{https://zenodo.org/}.
For PSR~J0740+6620 the \textit{NICER} observations have shown X-ray pulsations 
with two broad thermal X-ray peaks separated in phase by about $\sim$0.4.
In the case of PSR~J1614$-$2230, we have confirmed the X-ray pulsations 
originally reported by \citet{pancrazi12} and find one broad thermal 
X-ray peak in the phase diagram.
We show the fields of these pulsars in \xmmlong{} images and 
find that both pulsars are faint and nearby X-ray sources will contribute 
significantly to their respective \nicer{} concentrator count rates.

\acknowledgments
This work was supported in part by NASA through the \nicer{} mission and the 
Astrophysics Explorers Program. 
\nicer{} science team members at NRL are supported by NASA under Interagency
Agreement NNG200808A.
S.B.~was funded in part by NASA grants NNX17AC28G and 80NSSC20K0275. 
M.C.M. thanks the Radboud Excellence Initiative for supporting his 
stay at Radboud University. 
W.C.G.H. acknowledges support through grant 80NSSC20K0278 from NASA.
This research has made use of data and/or software provided by the 
High Energy Astrophysics Science Archive Research Center (HEASARC), 
which is a service of the Astrophysics Science Division at 
NASA/GSFC and the High Energy Astrophysics Division of the Smithsonian 
Astrophysical Observatory.  
We acknowledge use of NASA's Astrophysics Data System (ADS) bibliographic 
services and the ArXiv.
The authors thank the \xmmlong{} observatory, specifically the help of 
Felix Fuerst, for the Director's Discretionary Time observations of 
PSR~J0740+6620. 
\vspace{5mm}
\facilities{\nicer, \xmmlong}

\software{\texttt{astropy} \citep[ascl:1304.002]{astropy2013}, 
\texttt{PINT} (ascl:1902.007, \url{https://github.com/nanograv/pint}),
HEAsoft (ascl:1408.004, \url{https://heasarc.nasa.gov/lheasoft/})}, 
DS9 ( \citet{JoyeMandel2003}, ascl:0003.002),
\texttt{emcee} (ascl:1303.002, \url{https://github.com/dfm/emcee}), and 
the \xmmlong{} Scientific Analysis System (SAS, ascl:1404.004, 
\url{https://www.cosmos.esa.int/web/xmm-newton/sas}).

\bibliographystyle{aasjournal}
\bibliography{biblio}

\end{document}